# Anisotropic Characteristic Lengths of Colloidal Monolayers Near a Water-Air Interface

Na Li,[1] Wei Zhang, [2,a)] Zehui Jiang [3] and Wei Chen [1,b)]

*[1]State Key Laboratory of Physics and Department of Physics, Fudan University, Shanghai 200433, China*

*2School of Physical Science and Technology, China University of Mining and Technology, Xuzhou 221116, China*

*3 Department of Physics, Harbin Institute of Technology, Harbin 150001, China*

Near-interface colloidal monolayers have often been used as model systems for research on hydrodynamics in biophysics and microfluidic systems. Using optical microscopy and multiparticle tracking techniques, the correlated diffusion of particles is experimentally measured in colloidal monolayers near a water-air interface. It is found that the characteristic lengths $\chi_{\parallel}$ and $\chi_{\perp}$ of such a colloidal monolayer are anisotropic in these two perpendicular directions. The former ($\chi_{\parallel}$) is equal to the Saffman length of the monolayer and reflects the continuous nature of the system in the longitudinal direction. The latter ($\chi_{\perp}$) is a function of both the Saffman length and the radius of the colloids and reflects the discrete nature of the system in the transverse direction. This discovery demonstrates that the hydrodynamics intrinsically follow different rules in these two directions in this system.

The characterization of the viscoelastic properties of colloidal suspensions has long been the subject of fundamental research due to the ubiquity of such suspensions in biology and industry [1]. Microrheological techniques have been widely employed for such measurements due to their advantages in probing the local material response in systems such as porous media [2, 3], biological membranes [4] and microfluidic devices [5-9]. Research has shown

[a)] Author to whom correspondence should be addressed. Electronic mail: wzhangph@gmail.com
[b)] Author to whom correspondence should be addressed. Electronic mail: phchenwei@fudan.edu.cn

that the viscoelastic properties of a colloidal suspension are strongly affected by the confining boundaries [10-17]. Two-dimensional (2D) colloidal monolayers have traditionally been used to model the dynamic behaviors of proteins and other large molecules near a biomembrane [4, 14, 18, 19]. The hydrodynamic mechanisms in a colloidal monolayer differ from those in a 3D bulk liquid or a 2D liquid film [15, 20-22]. The mass of such a 2D monolayer is conserved within the monolayer, while momentum can propagate between the monolayer and the surrounding 3D liquid [15, 21]. To characterize the hydrodynamic interactions (HIs) between the particles in such a confined quasi-2D system, a characteristic length should be introduced. For a continuous two-fluid system, Saffman [18, 23] defined this characteristic length as $\lambda_s = \eta^{(s)}/\eta^{(b)}$, where $\eta^{(s)}$ is the viscosity of the liquid membrane and $\eta^{(b)}$ is the viscosity of the surrounding liquid. When the distance $r$ between two particles is much smaller than $\lambda_s$, the momentum is conserved in the 2D membrane. When the distance $r$ is much larger than $\lambda_s$, the momentum diffuses into the surrounding 3D liquid [18, 23-26].

In Saffman's model, the stress (momentum flux) in the membrane is spatially isotropic and decays logarithmically as $\sim \log(1/r)$ [18, 23] due to the conservation of momentum in the 2D liquid [20]. The characteristic length in Saffman's model is solely characterized by $\lambda_s$. This has been experimentally validated by the work of Weeks et al. [24] for a system consisting of a large-molecule membrane at a water-air interface. It has also been noted by Zhang [25] that the characteristic length in a particle monolayer at a water-air interface depends on both the particle size and the Saffman length. Previous work [15, 18, 21, 23-25] has mainly focused on the HIs in a liquid film suspended in a bulk liquid or at a liquid-liquid interface. By contrast, few experimental studies have been devoted to the dynamic features of particle monolayers close to a liquid-liquid interface, which will be distinctly different from those for a monolayer at the interface. Knowledge of the transition of the HIs from the 2D monolayer to the 3D bulk liquid is essential for understanding the role played by the interface in the HIs in the monolayer.

In this study, we report experimental investigations of the correlated diffusion of colloidal particles in a monolayer close to a water-air interface. It is found that the characteristic lengths in the longitudinal and transverse directions are different. In the longitudinal direction, the characteristic length is the Saffman length $\lambda_s$. In the transverse direction, the characteristic length is a function of $\lambda_s$ and the particle radius $a$. With these characteristic lengths, the curves describing the correlated diffusion of particles under different conditions can be collapsed into one master curve. Using these scaling factors, the viscosity of such a monolayer can be estimated, and the result is consistent with that obtained from one-particle measurements.

The experimental system is shown in Fig. 1(a). Samples of three kinds of silica spheres with radii of $a = 1.57\ \mu\text{m}$ (Si1), $1.0\ \mu\text{m}$ (Si2) and $0.6\ \mu\text{m}$ (Si3) were used in experiments. Following the experiment method in supplementary material (SM), the samples were prepared. The colloidal particles sank and formed a monolayer near the water-air interface under gravity. The separation between the center of the particle monolayer to the interface is denoted by $z$.The images of the particles monolayer, as seen in Fig. 1(b), were recorded by a CCD camera. The particle trajectories $\vec{s}(t)$ were obtained using a home-made particle tracking program.

The correlated diffusion reflects the response function of the HIs between two particles [24, 27, 28]. In terms of the particle displacement, the correlated diffusion coefficient is defined as

$$D_{\parallel,\perp}(r) = \frac{\langle \Delta s^i_{\parallel,\perp}(t,\tau)\Delta s^j_{\parallel,\perp}(t,\tau)\delta(r-R^{i,j}(t))\ \rangle_{i\neq j}}{2\tau}. \quad (1)$$

Here, $\Delta s^i_{\parallel,\perp}$ is the displacement of the $i$th particle in the longitudinal $(\parallel)$ or transverse $(\perp)$ direction during a time interval $\tau$, as shown in Fig. 1(c). The average $\langle\ \rangle_{i\neq j}$ is taken over all pairs consisting of the $i$th and $j$th particles with a separation distance $r$ for $i \neq j$. The correlated diffusion coefficients $D_{\parallel,\perp}(r)$ for sample Si1 are plotted in Fig. 2(a,b), where the results are normalized with respect to the single-particle diffusion $D_s(n)$ to eliminate the effect of the local viscosity. $D_s(n)$ is obtained as described in note 1 in SM. In Fig. 2(a,b), from bottom to top, the area fraction $n$ varies from 0.03 to 0.59, and the black squares and green crosses correspond to the smallest and largest $n$ values, respectively. The curves of $D_{\parallel,\perp}(r)$ for samples Si2 and Si3 are plotted in Fig. S2 in SM and exhibit behaviours similar to those of $D_{\parallel,\perp}(r)$ for sample Si1.

As shown in Fig. 2(a,b), the curves of the dimensionless correlated diffusion coefficients $D_{\parallel,\perp}(r/(2a))/D_s(n)$ are still rather diverse. This suggests that $D_s(n)$ and $a$ are not good scaling factors. A good scaling factor should cause all scaled curves to collapse to a single master curve after the dimensionless treatment [24, 29]. The scaling factor $D_s(n)$ used in Fig. 2(a,b) is the single-particle diffusion coefficient, which is influenced by two factors: the local viscosity experienced by a single particle in the dilute limit and the many-body effect of the particles. According to the work of Sickert [30] and Fischer [31], the single-particle diffusion coefficient can be written as

$$D_s(n) \cong \frac{k_B T}{\kappa^{(0)}\eta^{(\mathrm{b})}a + \kappa^{(1)}\eta^{(\mathrm{s})}}, \qquad (2)$$

where $k_B$ is the Boltzmann constant, $\eta^{(\mathrm{b})}$ is the viscosity of the bulk liquid, and $\eta^{(\mathrm{s})}$ is the viscosity of the particle monolayer. The dimensionless coefficients $\kappa^{(0)}$ and $\kappa^{(1)}$ can be calculated according to the method developed by Fischer [31], who suggests that both $\kappa^{(0)}$ and $\kappa^{(1)}$ are only functions of the separation $z/a$. The separation $z$ is calculated as described in the supplementary note 1. The calculated values of $z$, $\kappa^{(0)}$ and $\kappa^{(1)}$ for the three samples are listed in Table I.

Table I. Parameters of the three samples.

| Sample | $a$ ($\mu m$) | $z$ ($\mu m$) | $z/a$ | $\kappa^{(0)}$ | $\kappa^{(1)}$ |
|---|---|---|---|---|---|
| Si1 | 1.57 | 1.89 | 1.21 | 18.0 | 0.23 |
| Si2 | 1.00 | 1.26 | 1.26 | 18.1 | 0.22 |
| Si3 | 0.60 | 2.81 | 4.68 | 18.8 | 0.10 |

In equation (2), the denominator is divided into two terms, $\kappa^{(0)}\eta^{(\mathrm{b})}a$ and $\kappa^{(1)}\eta^{(\mathrm{s})}$. Hence, the diffusion coefficient $D_s(n)$ can also be rewritten as a sum of two terms [25]:

$$\frac{\mathbf{1}}{\boldsymbol{D_s(n)}} = \frac{\mathbf{1}}{\boldsymbol{D_s(0)}} + \frac{\mathbf{1}}{\boldsymbol{D'_s(n)}}. \qquad (3)$$

Based on equations (2) and (3), two effective diffusion coefficients can be defined: $D_s(0) \equiv k_B T/\kappa^{(0)}\eta^{(\mathrm{b})}a$ and $D'_s(n) \equiv k_B T/\kappa^{(1)}\eta^{(\mathrm{s})}$. $D_s(0)$ is the diffusion coefficient of a single particle in the monolayer in the dilute limit $(n \to 0)$, which can be obtained as in the supplementary note 1. $D_s'(n)$ is a function of $n$ because the viscosity $\eta^{(s)}$ stems from the HIs between particles in the monolayer. Since the measured correlated diffusion coefficients $D_{\parallel,\perp}(r)$ describes the His between particles, $D_s'(n)$ is a more suitable scaling factor for this scenario than $D_s(n)$ is.

However, the scaled correlated diffusion coefficients $\widetilde{D}_{\parallel}(r/2\text{a}) = D_{\parallel}/D_s'(n)$ and $\widetilde{D}_{\perp}(r/2\text{a}) = D_{\perp}/D_s'(n)$ for sample Si1 are still diverse, as shown in Fig. S3(a,b) in SM. All $\widetilde{D}_{\parallel,\perp}(r)$ curves collapse to a single master curve for each direction when the distance $r$ is replaced with a new scaled distance, $R_{\parallel,\perp} = r/\chi_{\parallel,\perp}(n)$ (see Fig. 2(c,d)). Here, $\chi_{\parallel,\perp}(n)$ are adjustable parameters and are defined as the characteristic lengths of the particle monolayer (see the supplementary note 2). The curves of $\widetilde{D}_{\parallel}(R_{\parallel})$ and $\widetilde{D}_{\perp}(R_{\perp})$ at large $n$ fall on the upper left side of the master curve, which indicates that the effective distances $R_{\parallel,\perp}$ between particle pairs are shortened in a large-$n$ particle monolayer and that the particle pairs are more strongly correlated with each other. The curves of $\widetilde{D}_{\parallel,\perp}(r/2\text{a})$ for samples Si2 and Si3 are shown in Fig. S4 and Fig. S5, and the master curves of $\widetilde{D}_{\parallel,\perp}(R_{\parallel,\perp})$ are plotted in Fig. S6 and Fig. S7 in SM. The existence of master curves for $\widetilde{D}_{\parallel,\perp}(R_{\parallel,\perp})$ indicates that the HIs in particle monolayers with different $n$ values obey the same response rule. Notably, the characteristic lengths $\chi_{\parallel,\perp}$ also depend on $n$. Based on the concept of the Saffman length, the values of the characteristic lengths $\chi_{\parallel,\perp}(n)$ of such a monolayer are determined by the viscosity of the system [18], as discussed below.

In the vicinity of a fluid-fluid interface, the mobility of particles and the HIs between particles will be different from those in the bulk, which show a complex dependence on the separation $z$ [13, 14, 32, 33]. In our experiments, $\widetilde{D}_{\parallel}$ and $\widetilde{D}_{\perp}$ indeed depend on the separation $z$ (Fig. 2(e,f)). The larger $z$ is, the weaker the influence of the water-air interface. The master curves of $\widetilde{D}_{\parallel}(R_{\parallel})$ and $\widetilde{D}_{\perp}(R_{\perp})$ with different $z$ each degenerate to a single curve when $\widetilde{D}_{\parallel}(R_{\parallel})$ and $\widetilde{D}_{\perp}(R_{\perp})$ are multiplied by a factor of $(z/a)^{2/3}$ (Fig. 2(e,f)). This degeneracy of $\widetilde{D}_{\parallel,\perp}$ by a factor of $(z/a)^{2/3}$ at $z > 0$ indicates that, with the exception of the boundary effect, no dynamic mechanisms are introduced into the system by the water-air interface.

According to equation (2), the viscosity of a particle monolayer can be estimated from the single-particle diffusion $D_s(n)$; the viscosity obtained in this way is denoted by $\eta^{(s1)}$. The viscosity of the particle monolayer is also directly related to the characteristic lengths $\chi_{\parallel,\perp}$ [18, 24, 25]. The viscosity estimated from $\chi_{\parallel,\perp}$, which is obtained from measurements of the correlated diffusion coefficients $D_{\parallel,\perp}(r)$ of particle pairs, is denoted by $\eta^{(s2)}$. The values of $\eta^{(s1)}$ and $\eta^{(s2)}$, although they are obtained in different ways, should agree with each other since they describe the same monolayer [24, 25]. It is found that the characteristic length $\chi_\parallel$ is identical in form to the Saffman length $\lambda_s$ [18, 23], the expression for which is written as follows:

$$\boldsymbol{\chi_\parallel = \lambda_s = \frac{\eta_\parallel^{(s2)}}{\eta^{(b)}}}, \qquad (4)$$

where the viscosity $\eta_\parallel^{(s2)}$ obtained from $\chi_\parallel$ is consistent with the value $\eta^{(s1)}$ obtained from equation (2), as shown in Fig. 3(a). However, the characteristic length $\chi_\perp$ in the transverse direction cannot be determined using the functional form given in equation (4) (see the supplementary note 3) because the viscosity $\eta_\perp^{(s2)}$ obtained from $\chi_\perp = \eta_\perp^{(s2)}/\eta^{(b)}$ is not equal to $\eta^{(s1)}$. The viscosity $\eta_\perp^{(s2)}$ follows the power-law relationship $\eta_\perp^{(s2)} \sim (\eta^{(s1)})^{2/3}$ when $\chi_\perp = \lambda_s$ is used (see Fig. 3(b)). Considering this power-law relationship, the dependence of the characteristic length $\chi_\perp$ on the Saffman length should be expressed as

$$\boldsymbol{\chi_\perp = \frac{1}{2} a \left(\frac{\lambda_s}{a}\right)^{2/3}}, \qquad (5)$$

for $\eta_\perp^{(s2)} = \eta^{(s1)}$ to be satisfied. The viscosity $\eta_\perp^{(s2)}$ that is obtained using equation (5) is plotted against $\eta^{(s1)}$ in Fig. 3(a). In addition, the viscosity $\eta_\parallel^{(s2)}$ should, in principle, be identical to $\eta_\perp^{(s2)}$ since the viscosity of the monolayer is isotropic. The viscosity data obtained in our experiments actually indicate that $\eta_\parallel^{(s2)} = \eta_\perp^{(s2)}$, as seen from Fig. S8(a) in SM. Hence, the viscosity of the particle monolayer as obtained from the two-particle correlated diffusion, $\eta^{(s2)}$, is found to be $\eta^{(s2)} = \eta_\parallel^{(s2)} = \eta_\perp^{(s2)}$. The dependence of $\eta^{(s2)}$ on $n$ is plotted in Fig. 3(c), where it is shown to follow the Krieger-Dougherty equation [34]. More details on this equation are provided in the supplementary note 4. The characteristic lengths in the two directions, $\chi_\perp$ and $\chi_\parallel$, are different, and the relationship between them is shown to take the form $\chi_{\perp}/a = (\chi_{\parallel}/a)^{2/3}/2$ in Fig. S8(b) in SM.

For a particle monolayer located just at the water-air interface (Fig. S9(a) in SM), the characteristic lengths in the longitudinal and transverse directions are identical to each other [24, 25]. In our experiments, however, the monolayer is located a short distance from the water-air interface (Fig. S9(b) in SM), and the characteristic lengths become different in the two directions. This phenomenon can be attributed to the boundary effect of the water-air interface. Figure 2(a) shows that $D_{\parallel}(\mathrm{r})$ decays with $r$ as $\sim 1/r$, in the longitudinal direction. This results from that the HIs response to a 3D-like shear stress in the bulk water and the thin-film water, which act as a kind of semi-3D system, duo to the momentum conservation in a 3D liquid [15, 21, 28, 35, 36]. While in Fig. 2(b), $D_{\perp}(\mathrm{r})$ decays as $\sim 1/r^2$ in the transverse direction, showing behavior that has been attributed to long-range compression and modeled as interactions of effective mass dipoles [6, 11]. This difference in the variation tendencies of the correlated diffusion coefficients in the longitudinal and transverse directions is universal among colloidal monolayers suspended in fluids [15, 21].

In our system, the characteristic length splits into a longitudinal characteristic length $\chi_{\parallel}$ and a transverse characteristic length $\chi_{\perp}$. The relationship between them is revealed to follow $2\chi_{\perp}/a = (\chi_{\parallel}/a)^{2/3}$. This relationship may be understood by analogy with the lubrication of a liquid film confined between two solid surfaces. The normal load capacity of the film depends on the form of hydrodynamic action that the film experiences [37]. In the case of squeezing action, the normal load capacity is $W_{\mathrm{squeeze}} = (w'/\eta^{(\mathrm{b})} u_{\mathrm{squeeze}})(h_{\mathrm{squeeze}}/l)^3$, where $w'$, $u_{\mathrm{squeeze}}$, $h_{\mathrm{squeeze}}$, and $l$ are the normal load per unit length, the squeezing velocity, the film thickness, and the length of the solid surface, respectively. Similarly, the expression for the load capacity for sliding action [37] is $\mathrm{W}_{\mathrm{sliding}} = (w'/(\eta^{(\mathrm{b})} u_{\mathrm{sliding}}))(h_{\mathrm{sliding}}/l)^2$, where $h_{\mathrm{sliding}}$ is the thickness of the liquid film and $u_{\mathrm{sliding}}$ is the sliding velocity. When these two load capacities become comparable (i.e., $W_{\mathrm{squeeze}} \sim \mathrm{W}_{\mathrm{sliding}}$ ) with $u_{\mathrm{squeeze}} = u_{\mathrm{sliding}}$ , the equation $h_{\mathrm{squeeze}}/l \sim (h_{\mathrm{sliding}}/l)^{2/3}$ is obtained, with a functional form similar to that of $2\chi_{\perp}/a = (\chi_{\parallel}/a)^{2/3}$. The analogy made here is based on the recognition that the squeezing and sliding actions for a confined film are equivalent to the compression and shear stress between the particles in a particle monolayer [15, 21]. Thus, the relationship between the two characteristic lengths of the particle monolayer has the same form as that of $h_{\mathrm{squeeze}}/l = (h_{\mathrm{sliding}}/l)^{2/3}$.

The form of the characteristic length $\chi_{\perp} = a(\lambda_s/a)^{2/3}/2$ can also be understood by analogy to lubrication theory. The squeezing force between two particles of radius $a$ with a separation distance $r$ in a liquid is $f = \xi(r)U$, where $U$ is the velocity at which one particle is approaching the other and $\xi(r)$ is the hydrodynamic friction coefficient. When the two particles are suspended in a 3D liquid with a separation distance $r_{3D}$, the friction coefficient is $\xi(r_{3D}) = (3/2)\pi\eta^{(\mathrm{b})}a(a/r_{3D})$ [38]. In a 2D system, such as two circular disks of radius $a$ with a separation distance $r_{2D}$ approaching each other in a thin liquid film [37], the friction coefficient of the squeezing force is $\xi(r_{2D}) = (3/2)\pi\eta_s(a/r_{2D})^{3/2}$. Here, the viscosity of the film, $\eta_s$, is equal to $\eta^{(\mathrm{b})}a$. By equating these two kinds of squeezing lubrication forces [i.e., $\xi(r_{2D}) = \xi(r_{3D})$], we find that $r_{2D} = a(r_{3D}/a)^{2/3}$, which is similar to $\chi_{\perp} = a(\lambda_s/a)^{2/3}/2$.

In this work, for a particle monolayer near a water-air interface, the correlated diffusion coefficients of the particles in the longitudinal and transverse directions are presented in the form of normalized functions $\widetilde{D}_{\parallel,\perp}(\tilde{R}_{\parallel,\perp})$. From such correlated diffusion measurements, one can obtain the characteristic lengths $\chi_{\parallel,\perp}$ of the particle monolayer, which are anisotropic in the longitudinal and transverse directions, satisfying the relation $2\,\chi_{\perp}/a = (\chi_{\parallel}/a)^{2/3}$. More specifically, the longitudinal characteristic length $\chi_{\parallel}$ is the Saffman length $\lambda_s$ of the particle monolayer, while the transverse characteristic length follows a power-law relationship with $\lambda_s$, as expressed by $\chi_{\perp} = a(\lambda_s/a)^{2/3}/2$. Using these characteristic lengths, the master curves of the correlated diffusion and viscosity of such particle monolayers can be obtained. Our experiments provide a set of reliable data that can be used for the further development of theoretical models for studying the dynamics of liquids near soft interfaces.

## Acknowledgments

This research is supported by the National Natural Science Foundation of China (Grant Nos. 11474054, 11774417 and 11604381), and the Natural Science Foundation of Jiangsu Province (Grant No. BK20160238).

# Figures

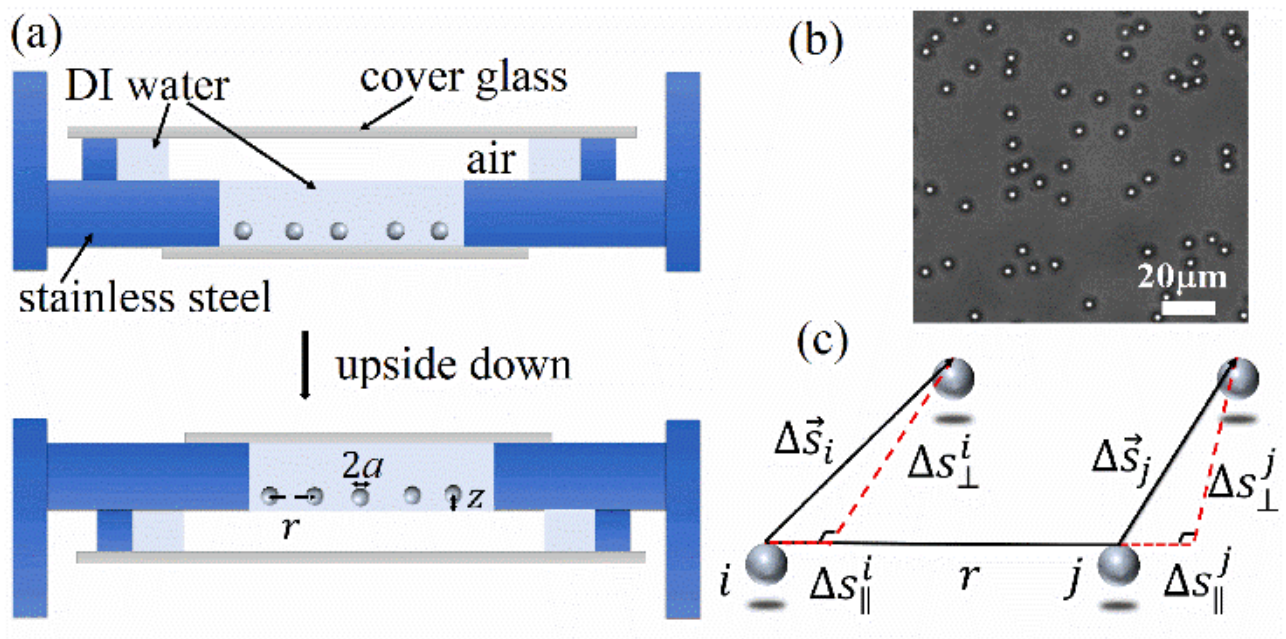

FIG. 1. (a) Schematic view of the system. (b) Optical microscope image of silica particles (a=1.57 μm) suspended near a water-air interface at an area fraction of $n = 0.04$. (c) The longitudinal displacement $\Delta s_\parallel$ and transverse displacement $\Delta s_\perp$ are the components of $\Delta\vec{s}_i(\tau)$ that are parallel and perpendicular, respectively, to the line connecting the centers of the two particles $i$ and $j$.

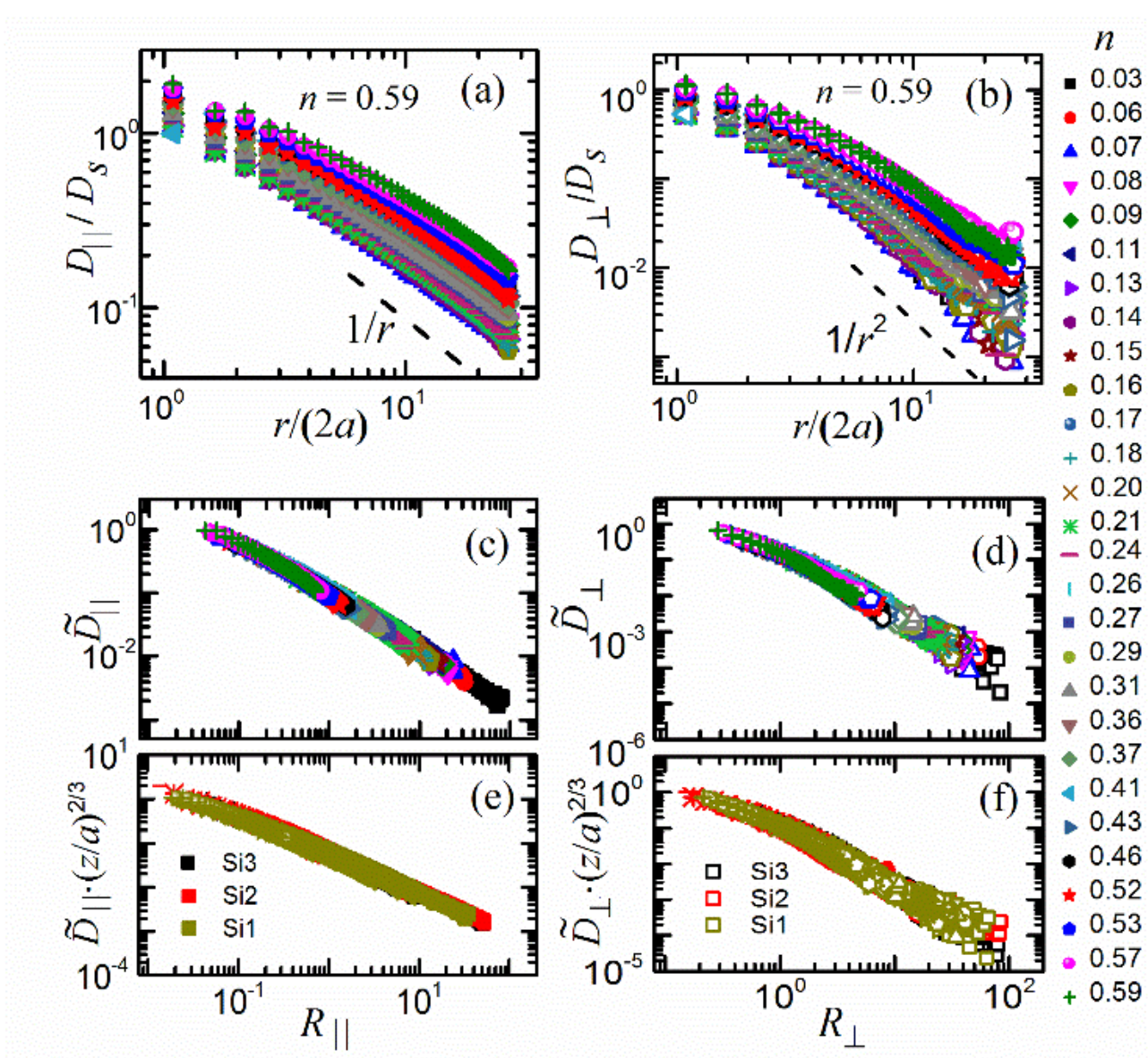

FIG. 2. Measured correlated diffusion coefficient $D_\parallel/D_s(n)$ (a) and $D_\perp/D_s(n)$ (b) as a function of the distance $r/(2a)$ for sample Si1. In (a) and (b), the different symbols represent measurements at different area fractions $n$, which vary from 0.03 (bottom) to 0.59 (top). The dashed lines corresponding to $\sim 1/r$ (a) and $\sim 1/r^2$ (b) are plotted as guides for the eye. (c) Scaled correlated diffusion coeffi-

cient $\widetilde{D}_{\parallel}$ as a function of the scaled distance $R_{\parallel}$ for sample Si1. (d) Scaled correlated diffusion coefficient $\widetilde{D}_{\perp}$ as a function of the scaled distance $R_{\perp}$ for sample Si1. (e) Universal master curve of $\widetilde{D}_{\parallel} \cdot (z/a)^{2/3}$ as a function of $R_{\parallel}$ for three samples. (f) Universal master curve of $\widetilde{D}_{\perp} \cdot (z/a)^{2/3}$ as a function of $R_{\perp}$ for three samples.

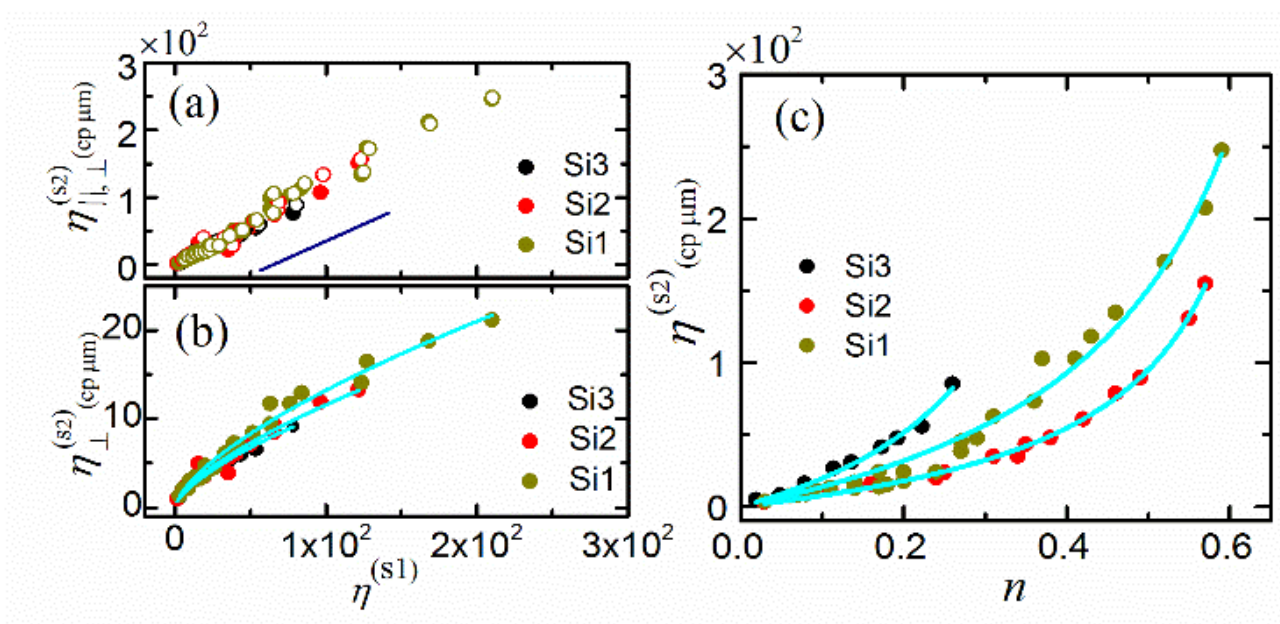

FIG. 3 (a) Plots of $\eta_{\parallel,\perp}^{(s2)}$ vs. $\eta^{(s1)}$ for three samples, where solid symbols represent $\eta_{\parallel}^{(s2)}$ and open symbols represent $\eta_{\perp}^{(s2)}$. The navy blue line is the guide for the eye, where the slope of the line is 1.0. (b) Comparison between $\eta_{\perp}^{(s2)}$ as calculated from $\chi_{\perp} = \lambda_s$ and $\eta^{(s1)}$ as obtained from equation (2). The cyan curves represent fits to $\eta_{\perp}^{(s2)} \sim (\eta^{(s1)})^{2/3}$. (c) Viscosities $\eta^{(s2)}$ as functions of the particle area fraction $n$. The cyan curves represent fits to the Krieger-Dougherty equation, which is presented in the supplementary note 4.

# Supplementary Material

**Experimental Method.**

Samples of three kinds of colloidal particles were purchased from Bangs Laboratories, namely, silica spheres with radii of $a = 1.57$ μm (Si1), 1.0 μm (Si2) and 0.6 μm (Si3). The particle samples were cleaned 8-10 times via centrifugation prior to use to scour off the surfactant in the solution. Then, the cleaned particles were suspended in deionized water (18.2 MΩ · cm) to form preparatory samples. The sample cell was made of stainless steel and had a structure similar to that in Ref. 1. The cell consisted of a solution tank at the bottom and an air tank at the top. The inner diameter of the solution tank was 8.3 mm. The depth of both was 0.8 mm. A preparatory sample was introduced into the solution tank, and the cell was sealed with a coverslip. Finally, the cell was placed upside down and allowed to remain undisturbed for 7-8 hours to allow the particles to settle down toward the water-air interface. A stable particle monolayer stays close to the water-air interface due to the interaction of the image charge of the particles. The surface tension of the water was sufficiently strong to retain the water in the top side of the cell. The experimental system is shown in Fig. 1(a). A microscope (Olympus X71 with 60x objectives) and a CCD camera (Prosilica GE1050, 1024*1024 pixels, 17 fps) were used to record images of the particles in the monolayer (Fig. 1(b)). The image resolution of the camera was 0.09 μm/pix. The particle trajectories $\vec{s}(t)$ were obtained using a homemade particle tracking program.

**Note 1. Diffusion coefficient of a single particle, $D_s(n)$.**

The single-particle self-diffusion coefficient $D_s(n)$ can be obtained from the mean square particle displacement $\langle \Delta\vec{s}_i^2(\tau) \rangle = 4D_s(n)\tau$, where $\Delta\vec{s}_i(\tau) = \vec{s}_i(t+\tau) - \vec{s}_i(t)$, $\tau$ is the lag time, and $n$ is the area fraction of the particles. Fig. S1 shows the normalized curves of $D_s(n)/D_0$ as functions of the particle concentration $n$ for the three samples, where $D_0$ is the diffusion coefficient for a single particle in the bulk water, $D_0 = k_B T/6\pi\eta a$. The value for each data point in Fig. S1 was obtained by averaging over more than $10^6$ particles. The solid lines in Fig. S1 illustrate the results of fitting the data to $D_s(n)/D_0 = \alpha(1 - \beta \cdot n - \gamma \cdot n^2)^2$. When $n \to 0$, $D_s(0)/D_0 = \alpha$, where $D_s(0)$ is the single-particle diffusion coefficient in the monolayer in the dilute limit. The fitted values of the parameters $\alpha$, $\beta$ and $\gamma$ for the three samples are given in Supplementary Table I. As seen from this table, a larger particle size corresponds to a larger value of $\alpha$ and a smaller value of $\beta$. Here, the value of the parameter $\alpha$ reflects the strength of the viscosity experienced by a single particle in the local environment in the dilute solution limit. A large $\alpha$ implies a low viscosity [3]. The value of the parameter $\beta$ represents the strength of the effective hydrodynamic interactions between two particles, excluding the effects of the local environment. A large $\beta$ reflects strong hydrodynamic interactions. The value of the parameter $\gamma$ represents the strength of the many-body effect among the particles in the monolayer. As shown in Supplementary Table I, a large $\alpha$ corresponds to a small separation $z/a$, which suggests that the local viscosity is stronger when the position of the particle monolayer in the water is farther from the water-air interface.

The separation $z$ is calculated according to equation (S1) [4, 5]:

$$\frac{D_s(0)}{D_0}=1+\frac{3}{16}\left(\frac{2\eta^{(\mathrm{b})}-3\eta^{(\mathrm{a})}}{\eta^{(\mathrm{b})}+\eta^{(\mathrm{a})}}\right)\left(\frac{a}{z}\right), \tag{S1}$$

where $\eta^{(\mathrm{b})}$ is the viscosity of the bulk water and $\eta^{(\mathrm{a})}$ is the viscosity of the air. The calculated values of $z$ are also shown in Supplementary Table 1.

**Note 2. Uniqueness of the scaling factor $\chi_{\parallel,\perp}$.**

There exist sets of values of $\chi_{\parallel,\perp}$ and their multiples such that all of these values can make $\widetilde{D}_{\parallel,\perp}$ collapse to single curves. However, there is only one special pair of $\chi_{\parallel,\perp}$ values for which the calculated viscosity $\eta^{(s2)}$ agrees with $\eta^{(s1)}$. This constraint allows one to determine a unique pair of $\chi_{\parallel,\perp}$ values with which to determine the positions of the master curves of $\widetilde{D}_{\parallel,\perp}$.

**Note 3. Relationships between the characteristic lengths and the viscosity of the particle monolayer.**

In a classical Saffman system, such as a biobilayer lipid film [6-8] suspended in a bulk liquid, the characteristic length of the film is $\lambda_s = \eta^{(s)}/\eta_b$ in both the longitudinal and transverse directions. In the present system, the characteristic length $\chi_{\parallel}$ in the longitudinal direction is also the Saffman length $\lambda_s$. The viscosity $\eta_{\parallel}^{(s2)}$ obtained from the relationship $\chi_{\parallel} = \lambda_s$ is consistent with the value of $\eta^{(s1)}$ obtained from the single-particle diffusion coefficient $D_s^{'}(n)$, as shown in Fig. 3(a) in the main text. However, the characteristic length in the transverse direction, $\chi_{\perp}$, cannot be taken to be equal to the Saffman length $\lambda_s$ because the viscosity $\eta_{\perp}^{(s2)}$ obtained from $\lambda_s$ differs from $\eta^{(s1)}$, as shown in Fig. 3(b) in the main text . In fact, $\eta_{\perp}^{(s2)}$ (as obtained from $\lambda_s$) and $\eta^{(s1)}$ exhibit a power-law relationship, $\eta_{\perp}^{(s2)} \sim (\eta^{(s1)})^{2/3}$, as shown in Fig. 3(b) in the main text. As suggested by this power-law relationship, the scaling factor $\chi_{\perp}$ should be written as $a(\lambda_s/a)^{2/3}/2$. The viscosity $\eta_{\perp}^{(s2)}$ obtained from $\chi_{\perp} = a(\lambda_s/a)^{2/3}/2$ agrees well with $\eta^{(s1)}$. The viscosity $\eta_{\perp}^{(s2)}$ also agrees with $\eta_{\parallel}^{(s2)}$, as shown in Fig. S7(a).

**Note 4. Dependence of the viscosity of the monolayer on the concentration.**

In the main text, Fig. 3(c) shows the monolayer viscosities $\eta^{(s2)}$ obtained from the correlated diffusion as a function of $n$ for the three samples. The dependence of $\eta^{(s2)}$ on $n$ is related to various properties of the sample, such as the colloidal particle size $a$ and the separation $z$. The solid cyan curves in Fig. 3(c) are fits to the Krieger-Dougherty equation [9]:

$$\eta^{(s2)} = \eta^{(s1)}(0)[(1 - \frac{n}{n_m})^{-[\eta]n_m} - 1], \qquad \text{(S2)}$$

where $\boldsymbol{\eta^{(s1)}(0)}=\boldsymbol{\eta_b a \kappa^{(0)}/\kappa^{(1)}}$ is the equivalent viscosity felt by a single particle in the particle monolayer in the dilute limit, which is affected by the confining boundary, and $\boldsymbol{\kappa^{(0)}}$ and $\boldsymbol{\kappa^{(1)}}$ are known coefficients [10]. In equation (S2), $\boldsymbol{n_m}$ is the packing fraction. When the concentration of the particle monolayer is close to $\boldsymbol{n_m}$, the viscosity of the monolayer becomes infinite. For hard spheres, the random packing fraction is $\boldsymbol{n_m \cong 0.84}$ in two dimensions [11, 12]. The intrinsic viscosity $[\boldsymbol{\eta}]$ is the only fitting parameter in equation (S1). The experimental data for the samples are well described by the solid cyan curves in Fig. 3(c) in the main text. The fitted values of $[\boldsymbol{\eta}]$ are listed in Supplementary Table I.

# Supplementary Figures

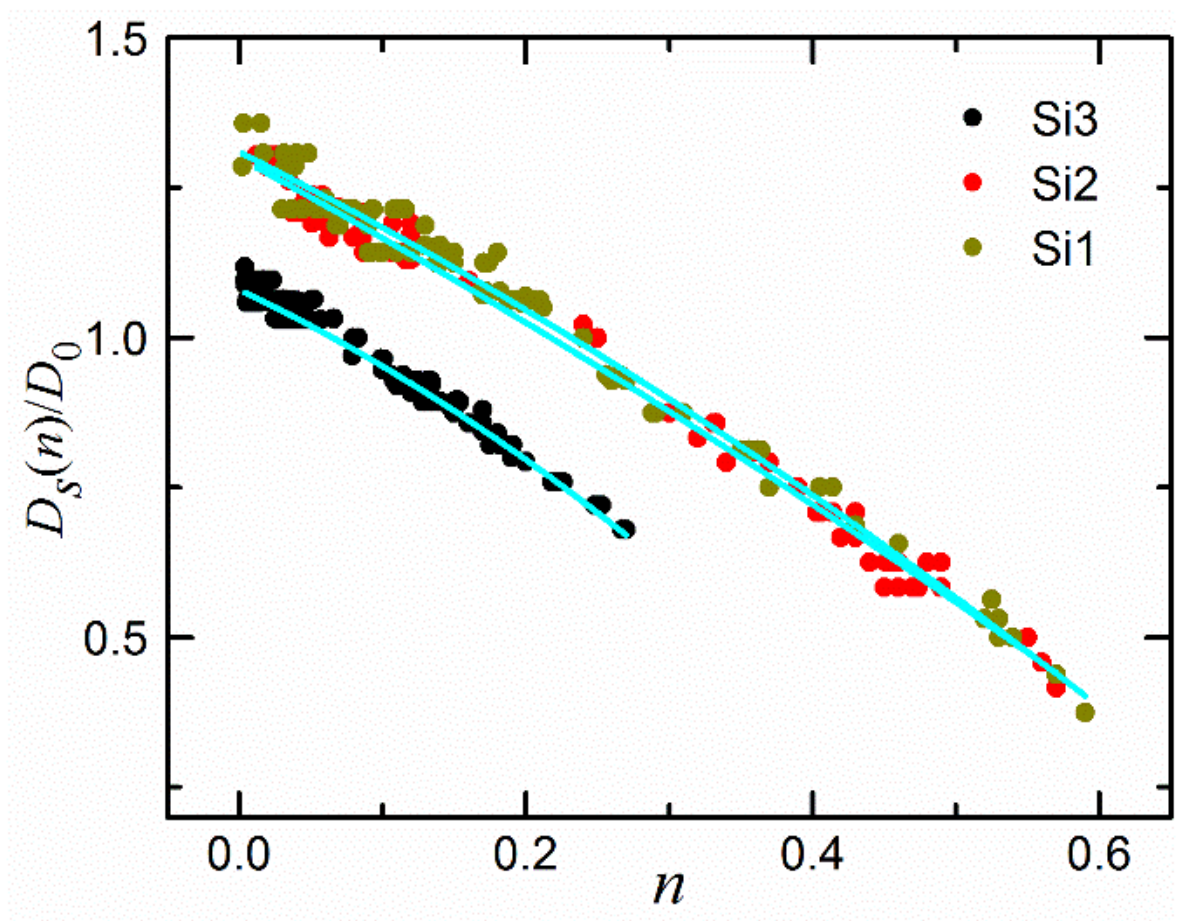

FIG. S1. Normalized self-diffusion coefficients of a single particle. The normalized self-diffusion coefficient $D_s(n)/D_0$ is plotted as a function of the area fraction $n$ for samples Si1, Si2 and Si3. The solid lines represent the second-order polynomial fits to the formula $D_s(n)/D_0 = \alpha(1 - \beta \cdot n - \gamma \cdot n^2)$ for each sample.

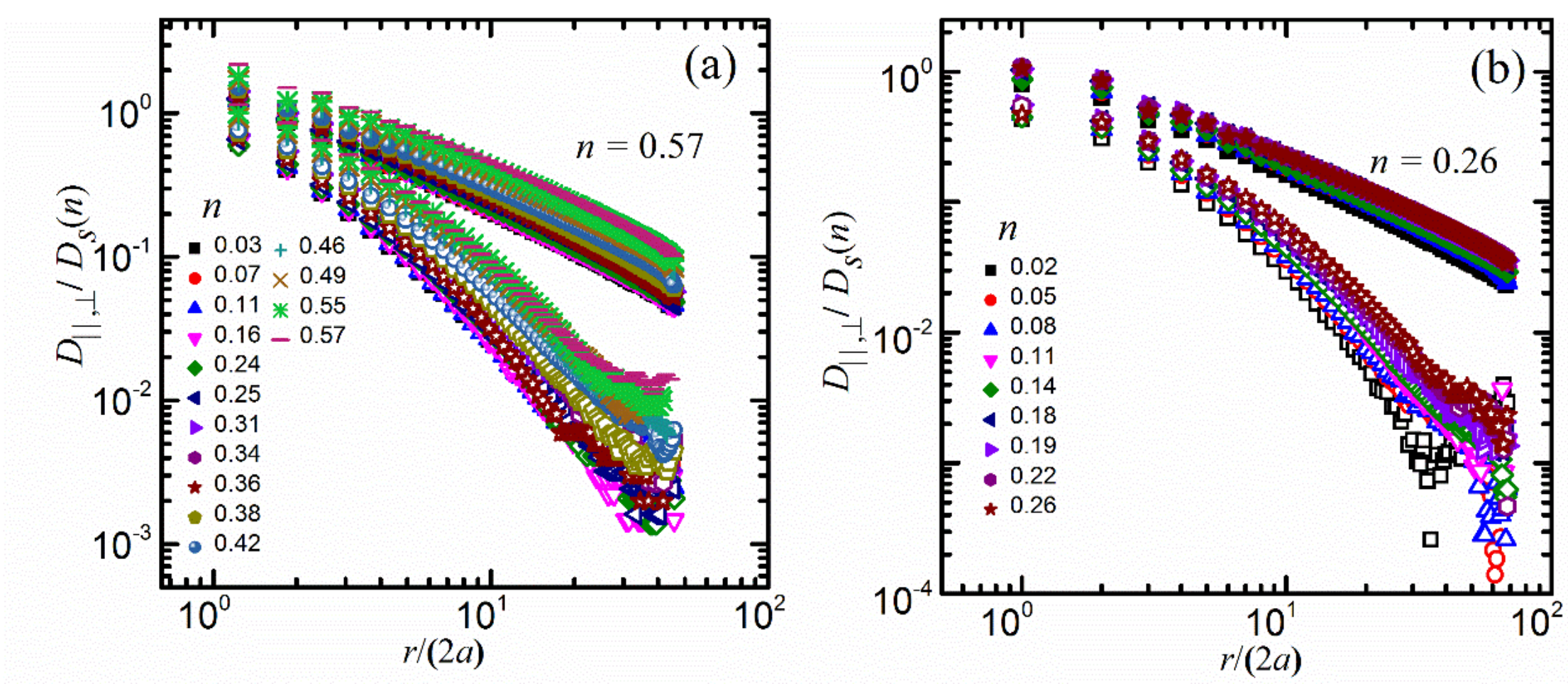

FIG. S2. Measured $D_{\parallel,\perp}/D_s$ values of samples Si2 and Si3 as functions of $r/(2a)$. (a) Measured correlated diffusion coefficients $D_{\parallel,\perp}/D_s$ of sample Si2 as functions of the distance $r/(2a)$ at different area fractions $n$ from 0.03 to 0.57. (b) Measured correlated diffusion coefficients $D_{\parallel,\perp}/D_s$ of sample Si3 as functions of $r/(2a)$ at different area fractions $n$ from 0.02 to 0.26. The different symbols represent $D_{\parallel,\perp}/D_s$ values measured at different concentrations $n$.

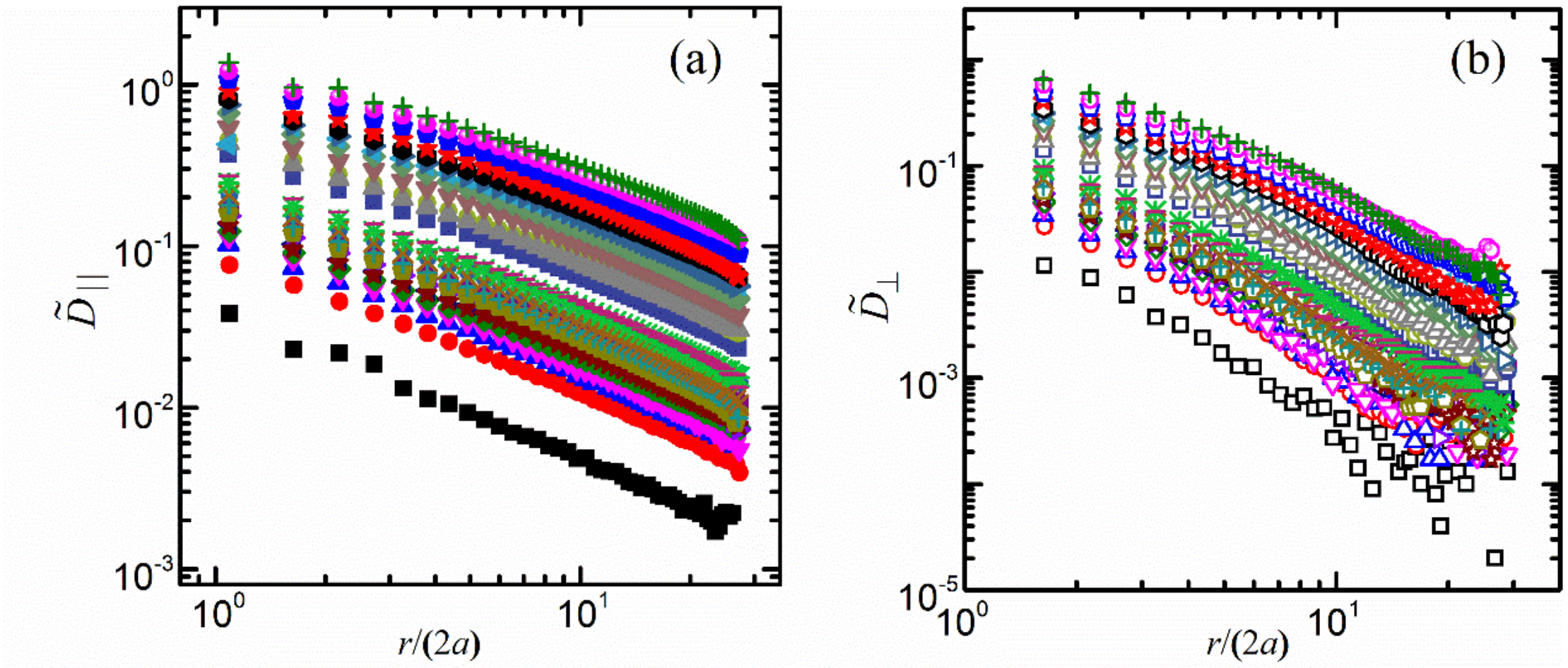

FIG. S3. Scaled correlated diffusion coefficient $\widetilde{D}_{\parallel}$ (a) and $\widetilde{D}_{\perp}$ (b) as a function of $r/(2a)$ for sample Si1. The symbols used are same as that in Fig. 2(a,b) in the main text. In (a) and (b), $\widetilde{D}_{\parallel,\perp}$ are obtained by $\widetilde{D}_{\parallel,\perp} = D_{\parallel,\perp}/D_s'(n)$.

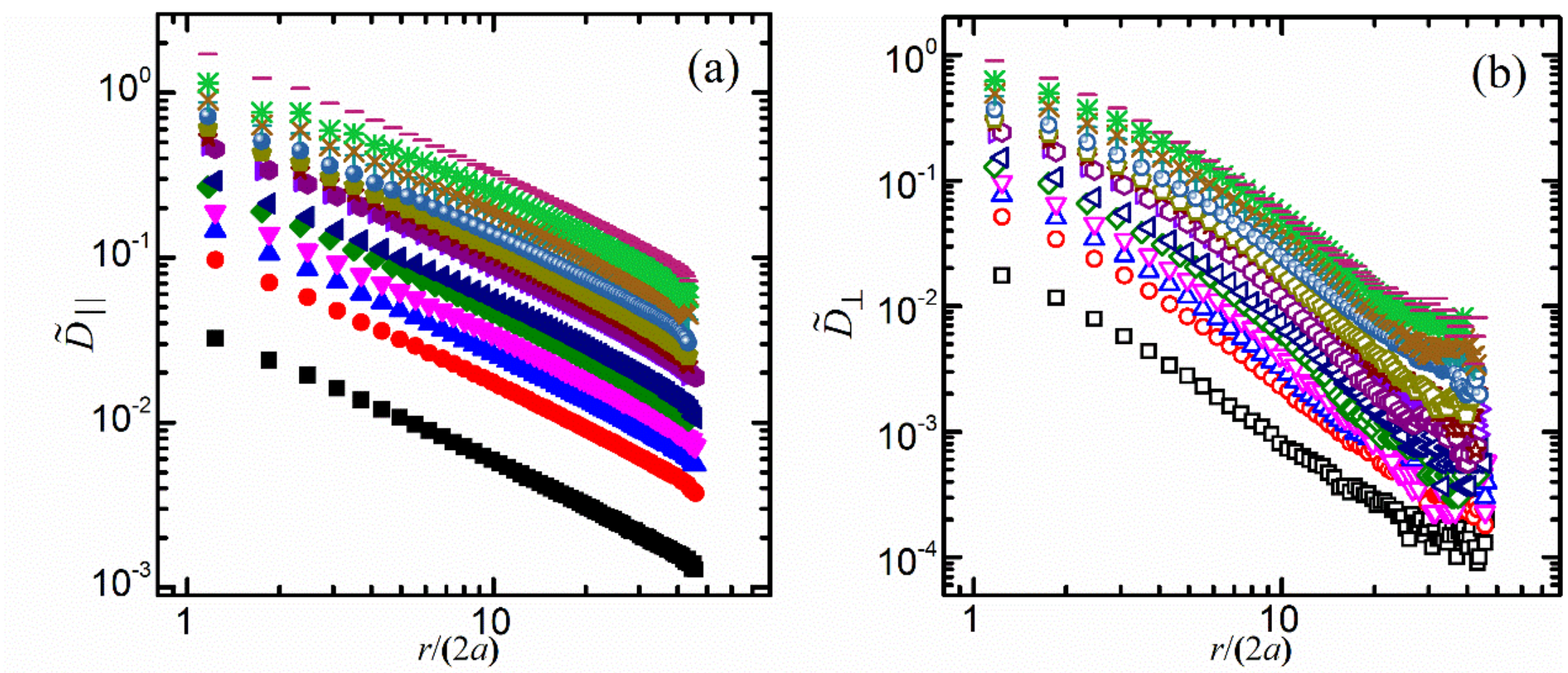

FIG. S4. Scaled correlated diffusion coefficient $\widetilde{D}_{\parallel}$ (a) and $\widetilde{D}_{\perp}$ (b) as a function of $r/(2a)$ for sample Si2. The symbols used are same as that in Fig. S2(a). In (a) and (b), $\widetilde{D}_{\parallel,\perp}$ are obtained by $\widetilde{D}_{\parallel,\perp} = D_{\parallel,\perp}/D_s'(n)$.

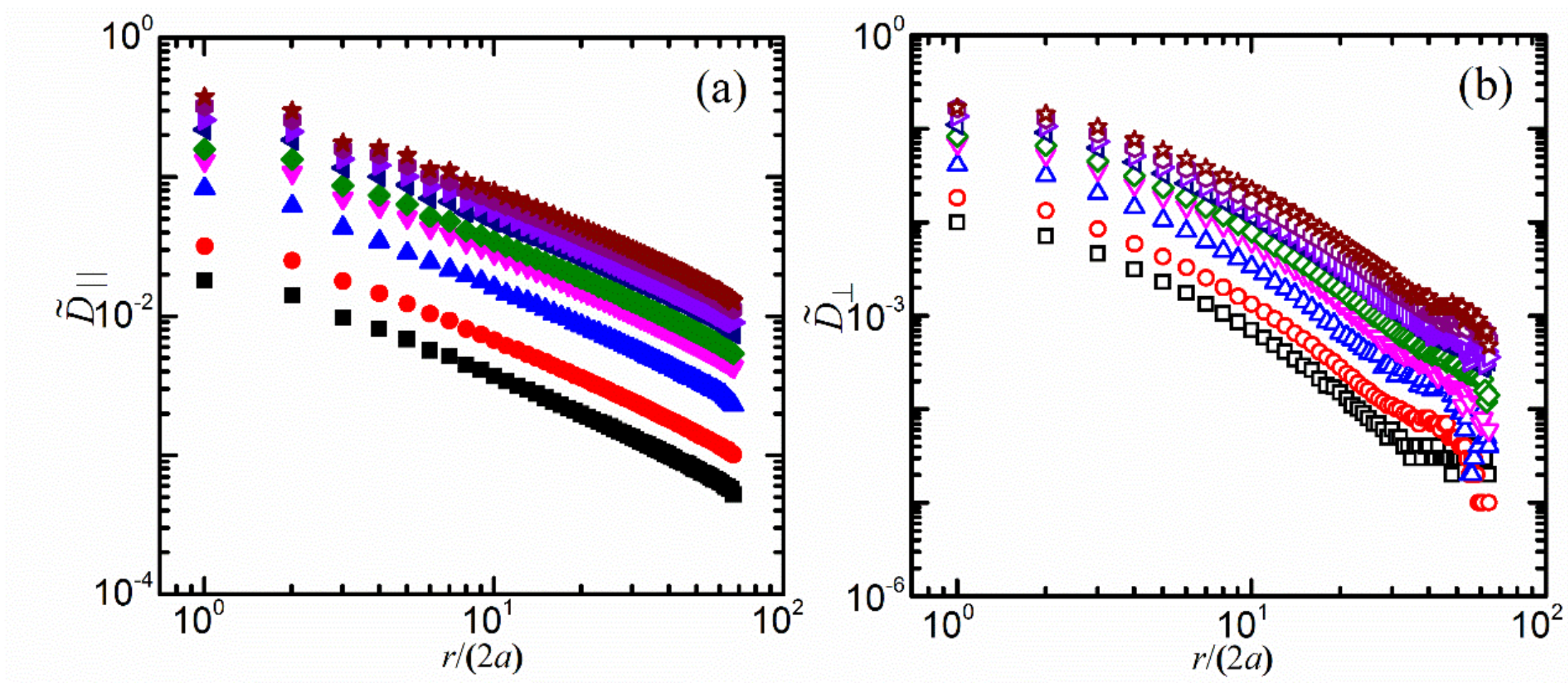

FIG. S5. Scaled correlated diffusion coefficient $\widetilde{D}_{\parallel}$ (a) and $\widetilde{D}_{\perp}$ (b) as a function of $r/(2a)$ for sample Si3. The symbols used are same as that in Fig. S2(b). In (a) and (b), $\widetilde{D}_{\parallel,\perp}$ are obtained by $\widetilde{D}_{\parallel,\perp} = D_{\parallel,\perp}/D'_s(n)$.

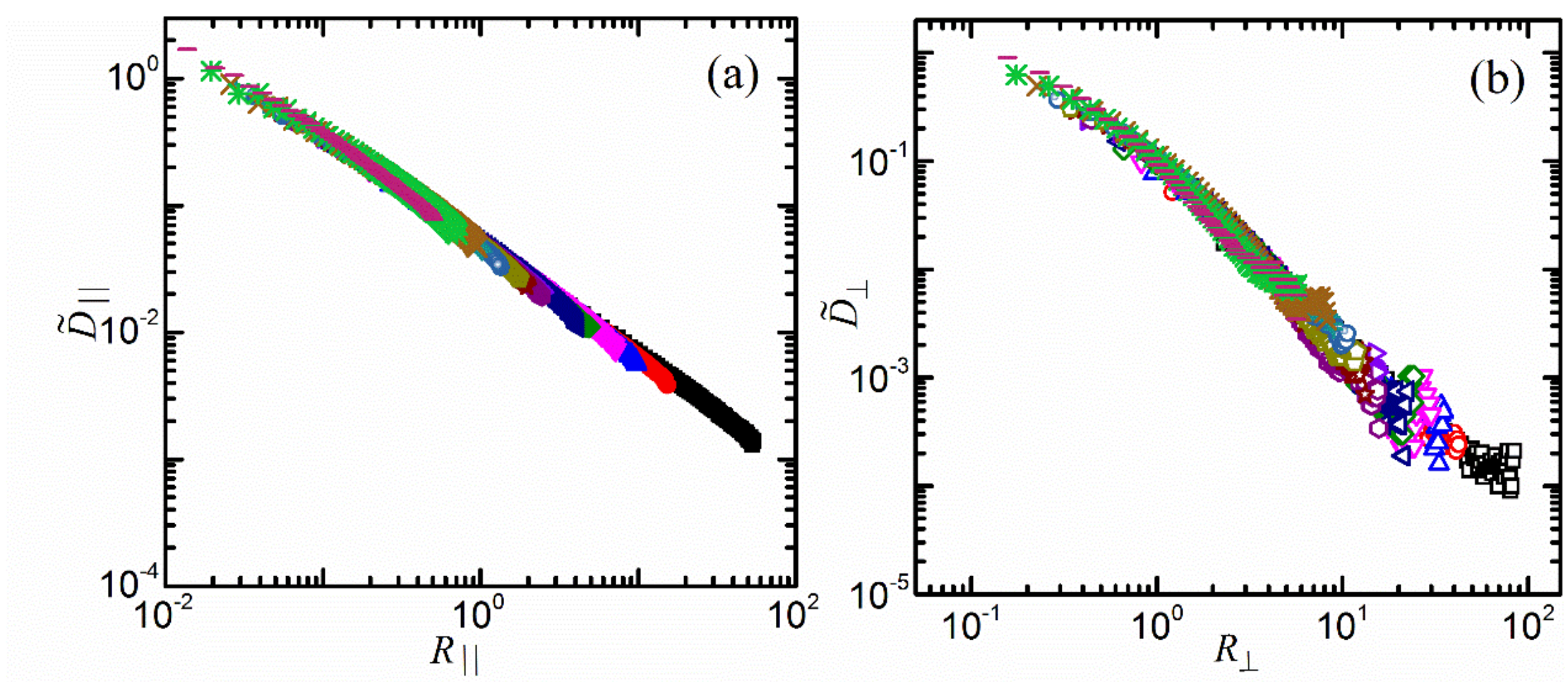

Fig. S6. (a) Scaled correlated diffusion coefficient $\widetilde{D}_{\parallel}$ as a function of the scaled distance $R_{\parallel}$ for sample Si2. (b) Scaled correlated diffusion coefficient $\widetilde{D}_{\perp}$ as a function of the scaled distance $R_{\perp}$ for sample Si2. The symbols used are same as that in Fig. S2(a). In (a) and (b), $R_{\parallel,\perp}$ is obtained by $R_{\parallel,\perp} = r/\chi_{\parallel,\perp}$.

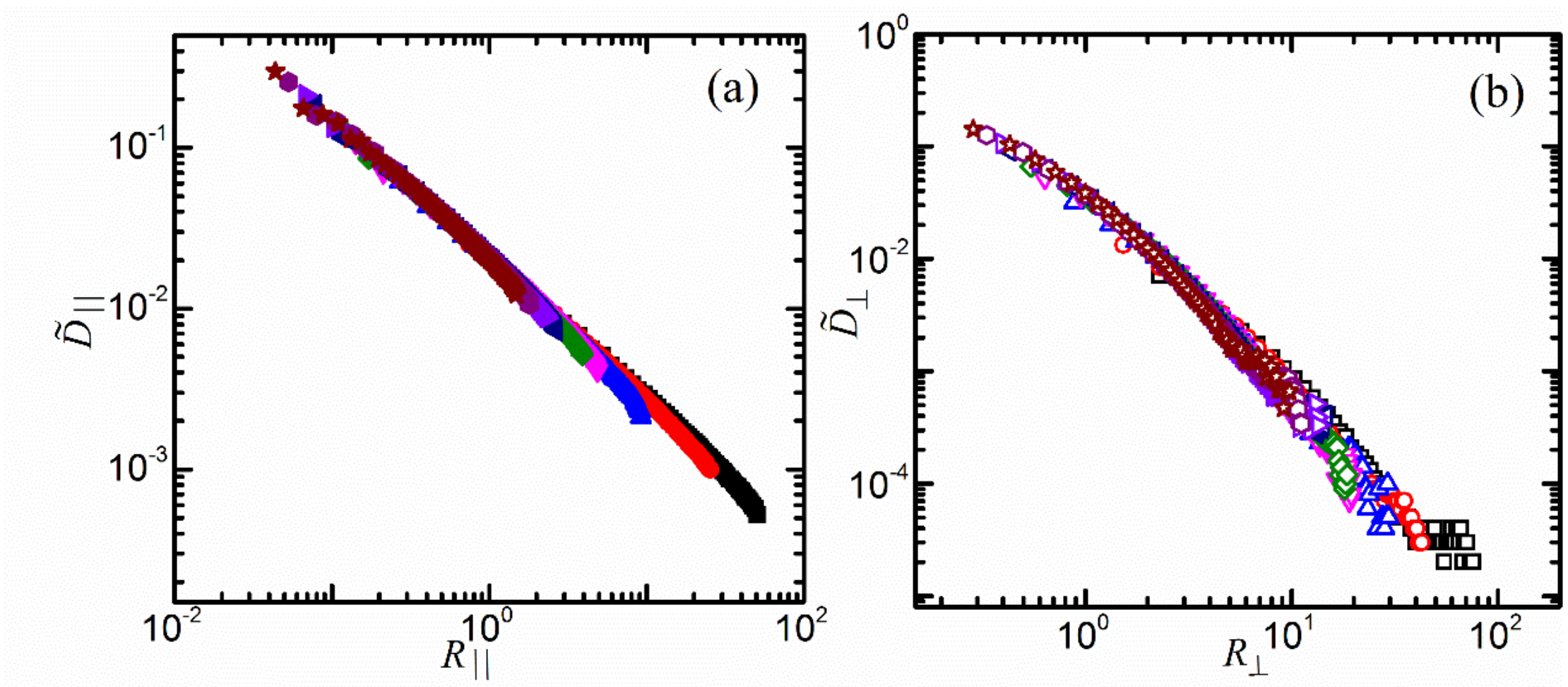

Fig. S7. (a) Scaled correlated diffusion coefficient $\widetilde{D}_{\parallel}$ as a function of the scaled distance $R_{\parallel}$ for sample Si3. (b) Scaled correlated diffusion coefficient $\widetilde{D}_{\perp}$ as a function of the scaled distance $R_{\perp}$ for sample Si3. The symbols used are same as that in Fig. S2(b). In (a) and (b), $R_{\parallel,\perp}$ is obtained by $R_{\parallel,\perp} = r/\chi_{\parallel,\perp}$.

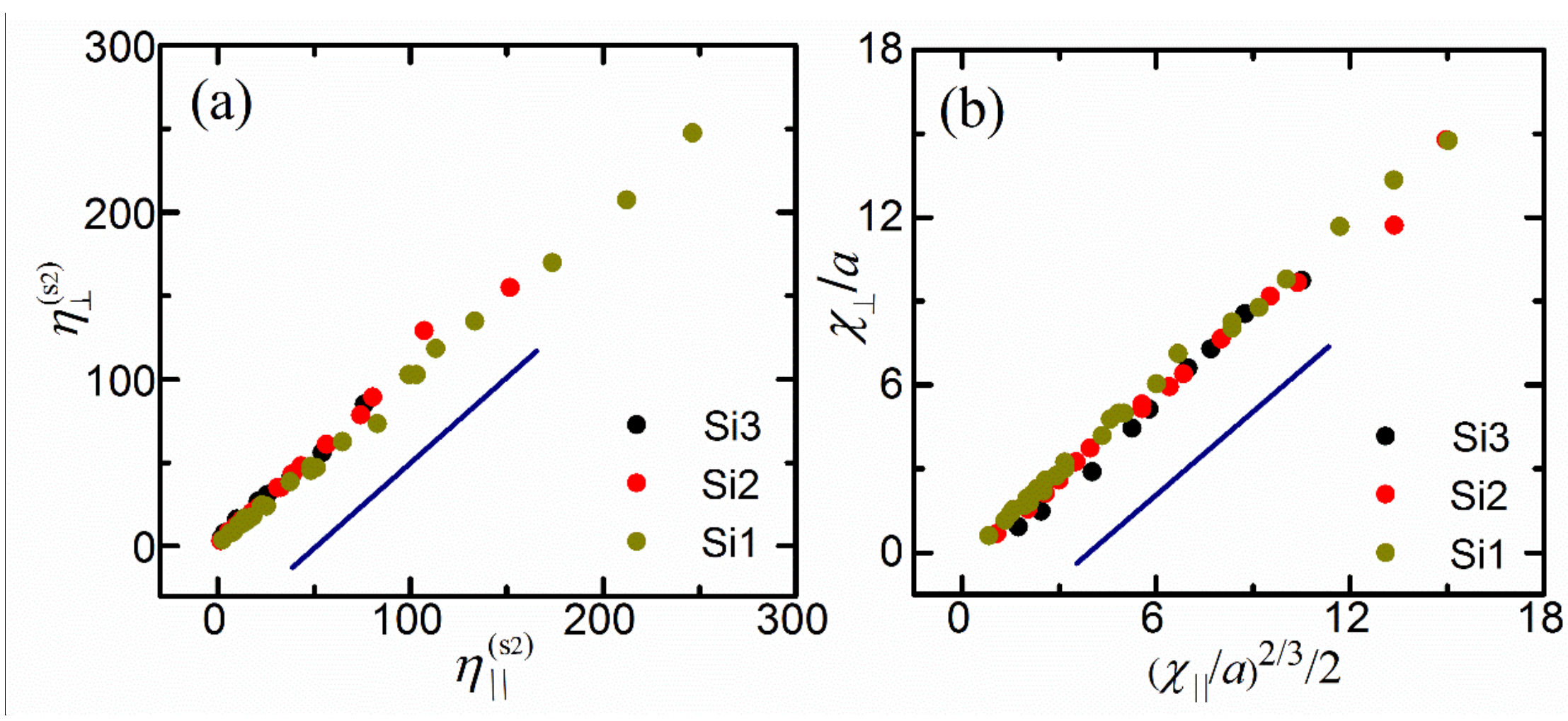

Fig. S8. (a) Plots of $\eta_{\perp}^{(s2)}$ vs. $\eta_{\parallel}^{(s2)}$ for three samples. (b) Plots of $\chi_{\perp}/a$ vs. $(\chi_{\parallel}/a)^{2/3}/2$ for three samples. The navy blue lines in (a) and (b) are guides for the eye, where the slope of each line is 1.0.

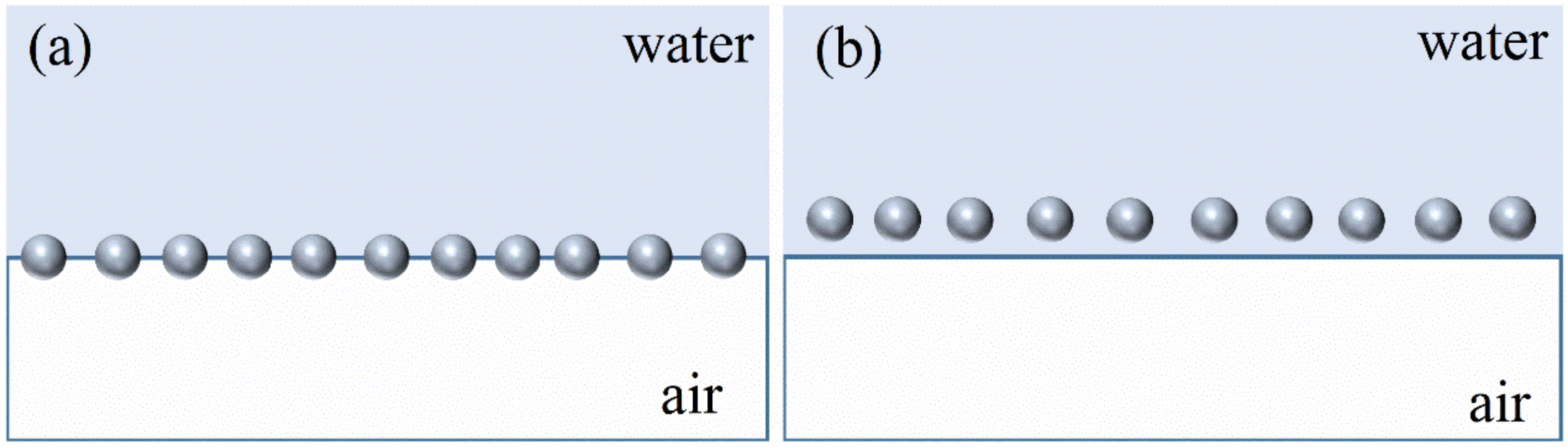

Fig. S9. Two kinds of colloidal systems. (a) The particle monolayer is at the water-air interface. (b) The particle monolayer is near the water-air interface.

# Supplementary Table

Supplementary Table I. Properties of samples Si1, Si2, and Si3: the particle radius $a$; the fitted parameters α, β, and γ of the polynomial function $D_s(n)/D_0 = \alpha(1 - \beta \cdot n - \gamma \cdot n^2)$; the separation $z$ as obtained from the experiments; and the intrinsic viscosity $[\eta]$ of the particle monolayer.

| **Sample** | $a(\mu m)$ | $\alpha$ | $\beta$ | $\gamma$ | $\alpha \cdot \beta$ | $z(\mu m)$ | $z/a$ | $[\eta]$ |
|---|---|---|---|---|---|---|---|---|
| Si1 | 1.57 | 1.31 | 0.92 | 0.43 | 1.21 | 1.89 | 1.21 | 1.15 |
| Si2 | 1.00 | 1.30 | 1.00 | 0.28 | 1.30 | 1.26 | 1.26 | 1.18 |
| Si3 | 0.60 | 1.08 | 1.05 | 1.31 | 1.13 | 2.81 | 4.68 | 1.89 |

# Supplementary references